\documentstyle[12pt]{article}
\begin{document}
\baselineskip 21pt plus .1pt minus .1pt
\def\r{\rightarrow}
\def\grav{\tilde\gamma\rightarrow\gamma\tilde G}
\def\neut{\tilde\gamma\rightarrow\gamma\nu_\mu}
\def\neue{\tilde\gamma\rightarrow\gamma\nu_e}
\def\neui{\tilde\gamma\rightarrow\gamma\nu_i}
\def\ben{\begin{equation}}
\def\een{\end{equation}}
\def\mml{m_{\tilde{e_L}}}
\def\mmr{m_{\tilde{e_R}}}
\pagestyle{empty}
\begin{flushright}
SINP/TNP/99-24\\ 
June 6, 1999
\end{flushright}
\vskip .1in
\begin{center}
\bf{R-parity Violating Radiative 
Photino Decay in Supersymmetric Models} 
\end{center}
\begin{center}
Ambar Ghosal
\end{center}
\vskip .1in
\begin{center}
Saha Institute of Nuclear Physics,\\
Theory Division,
Block AF, Sector 1, Salt Lake,\\
Calcutta 700 064, India\\
\end{center}
\vskip 24pt
\noindent
It has been shown that unless the 
tri-linear R-parity violating coupling 
$\lambda_{i33}$ ($i$ = 1, 2)  is small 
enough  
($\lambda_{i33}<$ ${10}^{-2}$ for MSSM and  
${10}^{-3}$ for GMSB model), 
the partial decay width of photino
decaying into 'photon + $\nu_{e,\mu}$', 
both in supergravity motivated 
(MSSM) and gauge mediated (GMSB) supersymmetric models 
are larger than the 
partial decay width of   
photino decaying into 'photon + goldstino' 
in R-parity conserving GMSB model 
including one loop supersymmetric QED correction. 
\vskip 1.0in
\noindent
PACS No. 12.60.Jv, 13.10.+q, 14.80.Ly \hfill
\vskip .1in 
\begin{flushleft}
E-mail: ambar@tnp.saha.ernet.in
\end{flushleft}
\noindent
(To appear in Phys. Rev. D)
\newpage
\pagestyle{plain}
\setcounter{page}{2}
Confirmation of neutrino oscillation 
by Superkamiokande 
experiment \cite{sk}  leads to the conclusion 
of non-zero neutrino mass. 
In Minimal Supersymmetric  
version of Standard Model  either 
Supergravity motivated 
(we refer it as MSSM) \cite{hb} or Gauge 
mediated (which we refer as GMSB) \cite{gm}, 
this feature of non-zero 
neutrino mass is realized through R- parity 
violation in the theory. 
Supersymmetric models with R parity violation 
opens up a plethora of new signals or can mimic 
the signals of R- parity conserving models. 
In the present work, we have computed such loop 
induced photino decays \cite{hall}  
$\neue$, $\neut$ via R- parity violation. 
The qualitative nature of both these processes 
are same and the quantitative difference arises 
due to the difference in respective R-parity 
violating couplings. Keeping this feature in view, 
in the following, we represent both the decays as    
$\neui$ (where $i$ = $e, \mu$) and the decay amplitude 
of both the processes will be evaluated just by 
replacing the respective R-parity violating coupling.
Furthermore, we have neglected the decay process 
$\tilde\gamma\rightarrow\gamma\nu_{\tau}$
as it is much suppressed compared 
to the other two processes. 
This is precisely because  
$\neui$ decays involve heaviest $\tau$ lepton in the loop 
whereas $\tilde\gamma\rightarrow\gamma\nu_{\tau}$
decay involves $e$ and $\mu$ leptons. The decay 
$\neui$ mimics the signal of $\grav$ in R-parity 
conserving GMSB model 
where $\tilde\gamma$ is the Next to Lightest 
Supersymmetric Particle (NLSP). Both these decay process 
, $\neui$ and $\grav$, give rise to the same final state 
 "$\gamma$ + ${\not{E}~~}$".
We have also considered one loop 
supersymmetric QED correction of 
the decay 
$\tilde\gamma$$\rightarrow$ $\gamma$$\tilde G$. 
There is not much enhancement 
in the partial decay width due to this 
correction and 
we find that the partial decay width 
of $\neui$ decay process   
is larger than the $\grav$ decay, 
unless the trilinear $\lambda_{i33}$ 
(where $i$ = 1, 2) coupling is too small. 
Furthermore, if R parity is violated , 
there will be possible three body photino decay 
($\tilde\gamma\r fff$) and it has been shown 
\cite{ka} that non-observation of such signal 
put a stringent constraint on the 
trilinear R-parity violating coupling $< 
{10}^{-5}$ , through the 
comparison between the partial decay width of 
$\grav$ with $\tilde\gamma\r fff$. 
A recent analysis \cite{sr} 
in this path has been done 
through the inclusion of bi-linear 
R-parity violating term and it has 
been shown that the 
branching ratio 
of $\tilde\chi_1^0\r\nu\gamma$ 
decay can have a maximum 
value of about 5 - 10$\%$. 
In the present work, we find that 
the tri-linear R-parity violating 
$\lambda_{i33}$ coupling alone give 
rise to a larger partial decay width of 
the decay process $\neui$ compared to the one loop 
supersymmetric QED corrected decay 
process $\grav$, 
unless the value of $\lambda_{i33}$ is too low. 
Thus, if R-parity is violated, 
ambiguity arises 
to interpret the observed signal 
"$\gamma +{\not{E}~~}$" 
or "$\gamma\gamma + {{\not{E}~~}}$ etc. 
\cite{cdf},\cite{ag} as a low energy 
signature of R-parity conserving GMSB model 
in an unambiguous way. 
Some other complementary signal in collider 
experiment should be needed which when 
taken into account with the 
"photon + missing energy " 
signal could lead us to confirm any 
of these models. 
Before going into the details, we like to mention 
the followings: First, 
although, in general, lightest neutralino 
$\tilde\chi_1^0$ is an admixture of the neutral 
gauginos and neutral Higgsinos, however, the 
present state of knowledge leads to the fact 
that the $\tilde\gamma$ component is dominated 
over the largest region of allowed parameter space
\cite{rob}.    
The relevant mixing factor arises due to 
general consideration of $\tilde\chi_1^0$ structure 
will modify 
equally all the decays discussed 
in the present work.
Second, 
we discard any photino-lepton-slepton off 
diagonal coupling in the present work.\hfill 
\vskip .1in  
To compute one loop supersymmetric 
QED correction to the decay of $\grav$ 
in R-parity conserving GMSB model,
we consider the following 
goldstino-lepton-slepton interaction 
Lagrangian \cite{fayet}\hfill
\ben
\it{L} = -i e_{gL}{\sqrt{2}}[{\bar{e_L}}
{\tilde{e_L}}\tilde{G} 
+ {\bar{\tilde G}}{\tilde{e_L}}^\star e_L] 
+ i e_{gR}{\sqrt 2}
[{\bar{e_R}}{\tilde{e_R}}{\tilde G} 
+ {\bar{\tilde G}}{\tilde{e_R}}^\star e_R]  
\een
where
\begin{eqnarray}
e_{gL} = \frac{m_{\tilde{e_L}}^2 - m_e^2}{d},
e_{gR} = \frac{m_{\tilde{e_R}}^2 - m_e^2}{d},
d = {\sqrt{\frac{3}{4\pi}}} M_{Susy}^2
\end{eqnarray}
In the above expressions $\mml$ , 
$\mmr$ are the masses of the 
left-slepton and right-slepton and 
$M_{Susy}$ is the supersymmetry breaking 
scale parametrized in terms of parameter $d$. 
In GMSB model, masses of left-slepton and 
right-slepton are wide apart primarily due to 
their different representation under 
SU(2) gauge group and since $\mml>> \mmr$  
we have discarded the contribution due to $\mml$. 
Furthermore, we ignored any 
non-degeneracy in right-slepton masses 
and $\mmr$ represents mass of the right-selectron. 
The one loop supersymmetric 
QED corrected diagrams of 
the decay 
$\tilde\gamma(q)\r$ $\gamma(p_2){\tilde G}(p_1)$ 
is generated due to slepton- lepton 
particles in the loop. The squark-quark induced loop 
diagrams are neglected 
since $m_{\tilde q}>> m_{\tilde l}$. 
Neglecting lepton masses as well compared to 
selectron mass, we obtain the following matrix element 
\ben
-i M_{loop} = i (\frac{2 e^2}{16 d \pi^2})\mmr^2 
A {\bar u}(p_1)\gamma^\rho u(q)\epsilon_\rho^\star
\een
where 
\ben
A = -\frac{3}{2} {\rm{ln}}(1 + p -2p^2) 
+ \frac{p}{18} + \frac{143}{60}p^2
\een
and $p = \frac{{m_{\tilde\gamma}}^2}{{\mmr}^2}$ 
where $m_{\tilde\gamma}$ is the mass of the photino.
It is to be noted that as $p\r 0$ , 
still there is a non-zero contribution to 
the loop correction due to the 
presence of the second term in the 
right-hand side of Eqn.(4), 
which shows non-decoupling effect 
of the above process. 
This is basically due to the proportionality 
of the coupling of the Goldstino-lepton-slepton 
term in the lagrangian with the slepton mass squared.  
\vskip .1in 
The relevant part of the 
Lagrangian required to calculate tree level 
$\tilde\gamma(q)\r\gamma(p_2){\tilde G}(p_1)$ 
is given by \cite{tony} 
\ben
{\it{L}} = \frac{1}{2d}\partial_\mu{\bar{\tilde\gamma}}
\gamma^\mu [\gamma^\nu , 
\gamma^\rho]\tilde G \partial_{\nu} A_\rho + h. c.
\een
 and the tree level matrix element comes out as 
\ben
-iM_{Tree} = i \frac{3 m_{\tilde\gamma}^2}{2d} 
{\bar u(p_1)}\gamma^\rho u(q) \epsilon_\rho^\star(p_2)
\een
The total matrix element 
$M_{total}$ ( = tree level + one loop)
of the decay process 
$\grav$ can 
be written as 
\ben
M_{total} = M_{tree}( 1 + \Delta) = 
\frac{3 m_{\tilde\gamma}^2}{2d}
(1 + \frac{e^2}{12\pi^2}\frac{m_{\tilde{e_R}^2}}
{m_{\tilde\gamma}^2}A){\bar u(p_1)}\gamma^\rho 
u(q)\epsilon_\rho^\star (p_2)
\een
where $\Delta$ = $\frac{M_{loop}}{M_{tree}}$ 
is the enhancement factor.
For a typical mass 
value of $m_{\tilde\gamma}$ = 80 GeV 
and $m_{\tilde e_R}$ = 100 GeV 
which are allowed in GMSB model, 
we found the enhancement in $M_{total}$ due to 
one loop  correction is 
$\Delta$$\sim$ 6$\times$ ${10}^{-3}$ 
for three generations of leptons. 
For higher values of photino and 
right-selectron masses the correction 
becomes more and more insignificant. 
Thus , we find that the enhancement due to the 
one loop 
supersymmetric QED correction of the 
decay $\grav$ is insignificant 
compared to its tree level decay mode.\hfill      
\vskip .1in
Next, we consider the one loop 
decay of $\neui$ 
in MSSM induced by the tri-linear 
R-parity violating $\lambda_{i33}$ 
coupling.  
The relevant diagrams 
are obtained by replacing 
goldstino field of the 
previous process by the 
$\nu_i$ field with 
R parity violating $\lambda_{i33}$ coupling, 
however, 
unlike the previous case, 
there is a chirality flip in the 
internal lepton(s) line(s) due 
to Yukawa type nature of the 
R-parity violating interactions, 
and therefore, we cannot neglect 
lepton mass in this case. 
We have considered 
heaviest $\tau$ lepton contribution only and 
as we have considered photino-lepton-slepton 
flavour diagonal coupling, the other particle 
circulating in the loop is $\tilde{\tau_R}$ 
. 
Furthermore, we have 
ignored any non-degeneracy between 
$m_{\tilde{\tau_L}}$ and 
$m_{\tilde{\tau_R}}$ and we have also 
ignored 
$\lambda^\prime$ coupling 
induced $d-\tilde d$ interactions 
by considering $m_{\tilde d}>> m_{\tilde \tau_R}$. 
\vskip .1in
We consider the following 
R-parity violating trilinear interaction, 
\ben 
{\it{L}}_{\rm\not{R_p}} = 
\frac{\lambda_{i33}}{2}[ 
\tilde{\tau_L}\nu_{iL}\bar{\tau_R} 
+{(\tilde\tau_R)}^\star\,{\bar{(\nu_{iL})^c}}\,\tau_L] + h.c.
\een
The squared matrix element of 
the process $\neui$ comes out as 
\ben
{|M|^2}_{MSSM} 
 = 16 Q^2 [ 2A^2 - {B_1}^2(A_1 + C)(B+ C) - 
2 A B_1(B + C)]
\een
where
\ben
Q = (\frac{\lambda_{i33}\alpha}{4\sqrt 2\pi})
(\frac{m_\tau}{m_{\tilde \tau}^2})
m_{\tilde\gamma}^2
\een
\ben
A = \frac{t}{t - 1}{\rm{ln}}t - {\rm{ln}}t -1
\een 
\ben
A_1 = \frac{2}{1 - t}(t{\rm{ln}}t + 1 - t) - 1 
+ \frac{2}{{(1-t)}^2}(\frac{t^2}{4} 
- \frac{1}{4} - \frac{t^2}{2}{\rm{ln}}t)
\een
\ben
B = \frac{t}{1-t}{\rm{ln}}t + 1 
\een
\ben
B_1 = \frac{3}{t - 1}
\een
\ben
C = \frac{1}{{(1-t)}^2}
[t(1-\frac{t}{2}){\rm{ln}}t 
+ (t - \frac{1}{4}) - \frac{3t^2}{4}]
\een
and $t = \frac{m_\tau^2}{m_{\tilde \tau}^2}$. 
Neglecting higher powers of $t$ , 
we obtain a simpler 
expression for ${|M|^2}_{MSSM}$ as 
\ben
{|M|^2}_{MSSM} 
 = 16 Q^2 [ 2{(1 + {\rm{ln}}t)}^2 
+ \frac{9}{2}{\rm{ln}}t + \frac{45}{16}]
\een
The partial decay width comes out as 
\ben
\Gamma_{\not R_p}^{MSSM} = 
\frac{1}{16\pi}{|M|^2}_{MSSM}
\frac{1}{m_{\tilde\gamma}}.
\een
The partial decay width $\Gamma(\grav)$ 
in R-parity conserving GMSB 
model at the tree level is 
 given by \cite{gam} 
\ben
{\Gamma(\grav)}^{GMSB} = 
\frac{m_{\tilde\gamma}^5}{6 M_{Susy}^4}
\een
and for the previous 
choice of photino mass and $M$ = 150 TeV , 
the partial decay width comes out 
as ${\Gamma(\grav)}^{GMSB}$$\sim$ $0.10\times 
{10}^{-11}$ whereas 
$\Gamma_{\not R_p}^{MSSM}$$\sim$ 
$0.17\times {10}^{-7}\times 
{\lambda_{i33}^2}$ for $m_{\tilde \tau}$ = 200 GeV , 
$m_{\tilde\gamma}$ = 100 GeV. 
Thus , unless 
$\lambda_{i33}$ is very small $(< {10}^{-2}),$ 
$\Gamma(\neui)_{{\not R_p}}^{MSSM}>$ 
$\Gamma(\grav)^{GMSB}$. Such a value of $\lambda_{i33}$ 
is well within the present upper bounds : 
$\lambda_{233}< 0.09(\frac{m_{\tilde \tau}}{100 {\rm{GeV}}})$, 
$\lambda_{133}< 0.24(\frac{m_{\tilde \tau}}{100 {\rm{GeV}}})$   
\cite{han}.
\vskip .1in
Similar result is also obtained 
in case of GMSB model including R-parity 
violation.  The squared matrix element 
in this case is given by 
\begin{eqnarray}
{|M|^2}_{GMSB} 
 = & 4 {(\frac{\lambda_{i33}\alpha}
{4\sqrt 2 \pi})}^2 t_1 
[ 2{(1 + {\rm{ln}}t_1)}^2 + \frac{9}{2}{\rm{ln}}t_1 
+ \frac{45}{16}]
\frac{m_{\tilde\gamma}^4}{m_{\tilde \tau_R}^2}\\ 
& + {\rm{terms}}\,\,\, {\rm{containing}}\,\,\, m_{\tilde\tau_L} 
\end{eqnarray} 
where $t_1 = \frac{ m_\tau^2}{m_{\tilde \tau_R}^2}$ 
and as before we have neglected higher powers 
of $t_1$. 
We can also neglect left-slepton 
contribution in the above 
expression since $m_{\tilde \tau_L}>> m_{\tilde \tau_R}$ 
in GMSB model. The partial decay width comes out as 
\ben
\Gamma_{\not R_p}^{GMSB} = 
\frac{1}{16\pi}{|M|^2}_{GMSB}
\frac{1}{m_{\tilde\gamma}}.
\een
For a typical choice of model 
parameters, $m_{\tilde\gamma}$ = 80 GeV , 
$m_{\tilde \tau_R}$ = 100 GeV we obtain,  
$\Gamma_{\not R_p}^{GMSB}$ 
= $0.21\times {10}^{-7}\times {\lambda_{i33}}^2$. 
Hence , as before , 
unless $\lambda_{i33}<{10}^{-3}$, 
the partial decay width of R-parity 
violating photino decay ($\neui$) in 
GMSB model is larger than 
the R- parity conserving 
photino deacy $(\grav)$ mode. 
\vskip .1in
In summary, 
we have calculated 
partial decay width of 
one loop  
radiative photino 
decay ($\neui$) (where $i = e, \mu$) both in 
MSSM as well as 
GMSB models due to tri-linear 
R-parity violating interactions. 
We have also computed one loop 
supersymmetric QED corrected  
amplitude of the decay process 
$\grav$ in R-parity conserving 
GMSB model. We found that 
for a typical choice of model parameters 
the enhancement due to this correction 
, $\Delta$(=$\frac{M_{loop}}{M_{tree}}$) 
is of the order of $6\times {10}^{-3}$
 for three 
generations of leptons. 
We have compared the one loop QED corrected   
partial decay width of the decay $\grav$ 
with the R-parity violating $\neui$ 
decay for both MSSM and 
GMSB models and we found  
that unless the tri-linear R-parity 
violating $\lambda_{i33}$ (where $i$ = 1, 2) coupling is small 
enough ($\lambda_{i33}< {10}^{-2}$ for 
MSSM and ${10}^{-3}$ for GMSB 
model), 
the partial decay width of this 
loop induced process is larger than the 
photino decay  
$\grav$ in R-parity conserving GMSB model. 
The upshot of this analysis leads 
to a crucial position to interpret 
the collider signal "photon + missing energy" 
as a signature of R-parity conserving GMSB 
model in an unambiguous way.     
\vskip .5in 
Author acknowledges Biswarup Mukhopadhyaya, 
Uma Mahanta, Debajyoti Choudhuri, 
Gautam Bhattacharya, Anirban Kundu 
 and Sourov Roy for many 
helpful comments and discussions.

\end{document}